\documentclass[aip,amsmath,amssymb,onecoloumn, reprint]{revtex4-2}
\usepackage{float}
\usepackage[english]{babel}
\usepackage{dcolumn}
\usepackage{bm}
\usepackage{fancyhdr}
\usepackage{sidecap}
\usepackage{multirow}

\usepackage[utf8]{inputenc}
\usepackage[T1]{fontenc}
\usepackage{mathptmx}
\usepackage{etoolbox}

\makeatletter
\def\@email#1#2{%
	\endgroup
	\patchcmd{\titleblock@produce}
	{\frontmatter@RRAPformat}
	{\frontmatter@RRAPformat{\produce@RRAP{*#1\href{mailto:#2}{#2}}}\frontmatter@RRAPformat}
	{}{}
}%

\usepackage[dvipsnames]{xcolor}
\usepackage{graphicx}
\usepackage{braket}
\usepackage{tikz}
\usepackage{color}
\usepackage{amsmath}
\usepackage[colorlinks=True, linkcolor=blue, filecolor=magenta, urlcolor=blue,citecolor=blue]{hyperref}

\makeatother
\begin{document}
\preprint{AIP/123-QED}
	
\title{Multi-probe detection of domain nucleation across the metal-insulator transition in VO$_2$}

	\author{Shubhankar Paul}
	\email{shubhp@iitk.ac.in}
	\affiliation{Department of Physics, Indian Institute of Technology Kanpur, Kanpur 208016, India}
	\affiliation{Toyota Riken–Kyoto University Research Center (TRiKUC), Kyoto 606-8501, Japan}	
	\author{Giordano Mattoni}
	\affiliation{Toyota Riken–Kyoto University Research Center (TRiKUC), Kyoto 606-8501, Japan}
	\author{Amitava Ghosh}
	\author{Pooja Kesarwani}
	\affiliation{Department of Physics, Indian Institute of Technology Kanpur, Kanpur 208016, India}
	\author{Dipak Sahu}
	\affiliation{Department of Physics, Indian Institute of Technology Kanpur, Kanpur 208016, India}
	\author{Monika Ahlawat}
	\affiliation{Department of Chemistry, Indian Institute of Technology Kanpur, Kanpur 208016, India}
	\author{Ashok P}
	\author{Amit Verma}
	\affiliation{Department of Electrical Engineering, Indian Institute of Technology Kanpur,Kanpur 208016, India}
	\author{Vishal Govind Rao}
	\affiliation{Department of Chemistry, Indian Institute of Technology Kanpur, Kanpur 208016, India}
	\author{Chanchal Sow}
	\email{chanchal@iitk.ac.in}
	\affiliation{Department of Physics, Indian Institute of Technology Kanpur, Kanpur 208016, India}

	\date{\today}
	\begin{abstract}
Electronic and structural degrees of freedom are often intimately coupled in strongly correlated systems, which result in intriguing macroscopic and microscopic phenomena. Using the well-studied material VO$_2$ as a prototype, here we explore the domain distribution across the metal–insulator transition (MIT). We use macroscopic as well as microscopic techniques, such as first-order reversal curve (FORC) and infrared imaging, to probe the domain distributions across the MIT. This study compares MIT in thin films of VO$_2$ with different grain sizes grown by pulsed laser deposition and dc sputtering. We explore the relation between the nature of the FORC distribution and
the corresponding thermal hysteresis due to interactions between the supercooled metallic domains and surrounding insulating matrix. Our multi-probe study with quantitative analysis provides a correlation between the growth, domain interaction, and domain nucleation process in MIT.
 
\end{abstract}
\maketitle

The metal-insulator transition (MIT) driven by the self-organization of electrons in real space is often described using percolation theory \cite{essam1980percolation,choi1996mid,qazilbash2007mott, morin1959oxides,mott1968metal}. MIT pertains to a  drastic change in electrical conductivity of a material in response to external stimuli, such as heat \cite{morin1959oxides}, chemical doping \cite{majid2018insulator}, strain/pressure \cite{jian2017roles,bai2015pressure,arcangeletti2007evidence,muraoka2014persistent, yang2015surface, muraoka2002metal}, electric fields \cite{lee2008electrically}, or light \cite{morrison2014photoinduced,qazilbash2007mott}. This study is focused on VO$_2$, a key material of interest due to its first-order MIT at $T_\mathrm{MIT}$ $\sim$ 340 K accompanied by a structural change from the monoclinic insulating phase (M1: P2$_1$/$c$) to rutile metallic phase (R: P4$_2$/$mnm$) on heating \cite{majid2018insulator,morin1959oxides,jian2017roles,bai2015pressure,arcangeletti2007evidence,muraoka2014persistent, yang2015surface, muraoka2002metal,mott1968metal,lee2008electrically,morrison2014photoinduced,qazilbash2007mott,muraoka2014persistent, yang2015surface, muraoka2002metal,schwingenschlogl2004vanadium}. The feature of externally controlled MIT near room temperature makes VO$_2$ special for various technological applications such as optical memory, Mott memory \cite{zhou2015mott,kim2004mechanism}, Mott FET \cite{ruzmetov2010three}, room-temperature bolometers \cite{CHEN2001212}, and UV detectors \cite{li2022flexible}. Thermal hysteresis at the MIT of VO$_2$ depends on film thickness \cite{suh2004semiconductor}, strain, growth temperature \cite{youn2004growth}, grain size \cite{brassard2005grain}, substrate \cite{nakano2015distinct}, distribution of local domain, and non-stoichiometric composition (such as oxygen vacancies, vanadium oxidation state, etc. \cite{kim1994pulsed,jeong2013suppression})

Recent studies on nanocrystals of VO$_2$ using transmission electron microscopy with femtosecond time resolution revealed that the MIT is strongly influenced by the local strain distribution \cite{kim2023femtosecond}. The sharpness of the MIT in VO$_2$ depends on the growth process. Various methods \cite{nakano2013infrared,kim1993finite,zhang2018near,kim1994pulsed,kumar2004characteristics,lu1993preparation,zhang2016characterization,leroy2012structural} have been used to grow VO$_2$ thin films to achieve sharp hysteresis loop and a significant resistance drop. However, phase inhomogeneity caused by non-stoichiometry and interfacial strain provide an extra challenge to characterize the physical properties. 

Due to the complex nature of the MIT in VO$_2$, a detailed exploration of hysteresis behavior and its microscopic origin is necessary, especially in the context of domain nucleation and its interaction with the surrounding medium. These interactions play a pivotal role throughout the MIT transition. First-order reversal curve (FORC) technique is a powerful method for studying the hysteresis for various systems. Ramírez $et$ $al$. \cite{ramirez2009first} used FORC method extensively in the study of VO$_2$ systems, providing information regarding persistent of metallic domains at insulating region. These supercooled metallic domain can not affect the FORC distribution unless these domains interact with the surrounding insulating matrix \cite{ramirez2009first,kim2022nanoscale,sohn2015fractal,gu2022random}.Our study combines FORC and IR measurements for two types of samples with distinct grain sizes, and shows how the material properties strongly influence the nucleation behavior. In larger-grain samples, we observe a dominant single conduction channel with symmetric thermal hysteresis, whereas in smaller-grain samples, we find multiple conduction channels and an asymmetric hysteresis, consistent with the presence of supercooled metallic domains. Our analysis shows the correlation between material properties determined by different growth conditions and nucleation of metallic/insulating phases at the electronic phase transition of strongly correlated materials.

VO$_2$ thin films of nearly identical thickness ($\sim$ 100 nm) were grown on $c$-plane Al$_2$O$_3$ substrates using two different techniques: pulsed laser deposition (PLD), denoted as P-VO$_2$, and DC magnetron sputtering, denoted as S-VO$_2$. The P-VO$_2$ film was grown using a KrF excimer laser (248 nm,  2 J/cm$^2$, 5 Hz) at 550 $^{\circ}$C in oxygen environment of 0.03 mbar pressure, followed by cooling at 2.5 $^{\circ}$C/min. The S-VO$_2$ films was grown using an atmospheric pressure thermal oxidation process (APTO) \cite{ashok2020vanadium}. In this case, the vanadium film was oxidized in an atmospheric open-air environment at 450 $^{\circ}$C for a duration of 20 s, followed by quenching at the rate of 110 $^{\circ}$C/s. The crystal structure of VO$_2$ thin films was examined by X-ray diffraction (XRD) using PANanalytical, model Empyrean. The surface morphology of the films was characterized by atomic force microscopy (AFM) using Oxford instrument model MFP3D Origin (Probe type: PPP-NCLAu-10) in contact mode. Electrical transport measurements were performed using a commercial physical property measurement system (Quantum Design, PPMS 5000).
The resistivity was measured using a standard PPMS resistance bridge (Quantum design) operated in DC delta mode, where the current polarity is periodically reversed. Measurements were performed using the standard four-probe configuration, with the outer electrodes serving as current leads and the inner electrodes used for voltage detection. Silver epoxy (EPO-TEK, H20E) was used to form robust electrical contacts with minimal contact resistance \cite{paul2025growth}.
Thermal images were taken using an IR camera (Avio, InfRec R500, spatial resolution 21 $\mu$m, spectral range 8–14 $\mu$m). A Peltier cooling stage (Ampére, UT4070) with temperature controller (UTC-200A) was used to capture the IR image at various temperatures \cite{mattoni2020role}. The films were grown along the $b$-axis (supplementary Fig.\textcolor{blue}{S1}), with P-VO$_2$ closely matching the bulk lattice parameter (4.53 \AA) \cite{yang2015surface}, while S-VO$_2$ shows a shorter $b$-axis (4.41 \AA). Rocking curves around the (020) peak (Fig.\textcolor{blue}{S1}) show sharp reflections with FWHM of 0.1 $^{\circ}$ (P-VO$_2$) and 0.2$^{\circ}$ (S-VO$_2$), indicating low mosaicity. The RMS surface roughness for P-VO$_2$ is smaller ($\sim$ 0.53 nm) than for S-VO$_2$ ($\sim$ 4.37 nm) suggesting a possible layer-by-layer growth with tiny particulates (Fig.{S1}) \cite{jalili2010growth}.
The grain size, estimated using the Scherrer formula\cite{zhao2021low} (D = K$\lambda$ / $\beta$cos$\theta$, where K, $\theta$ and $\beta$ are the Scherrer constant, Bragg angle, and FWHM of the diffraction peak, respectively), is $\sim$ 40 nm and 20 nm for P-VO$_2$ and S‑VO$_2$, respectively. The smaller grains in S-VO$_2$ introduce more surface states and grain-boundary strain, which possibly broaden the metal-insulator transition \cite{zhao2021low,radue2013effect}.

\begin{figure}[b]
	\begin{center}
		\includegraphics[scale=0.37]{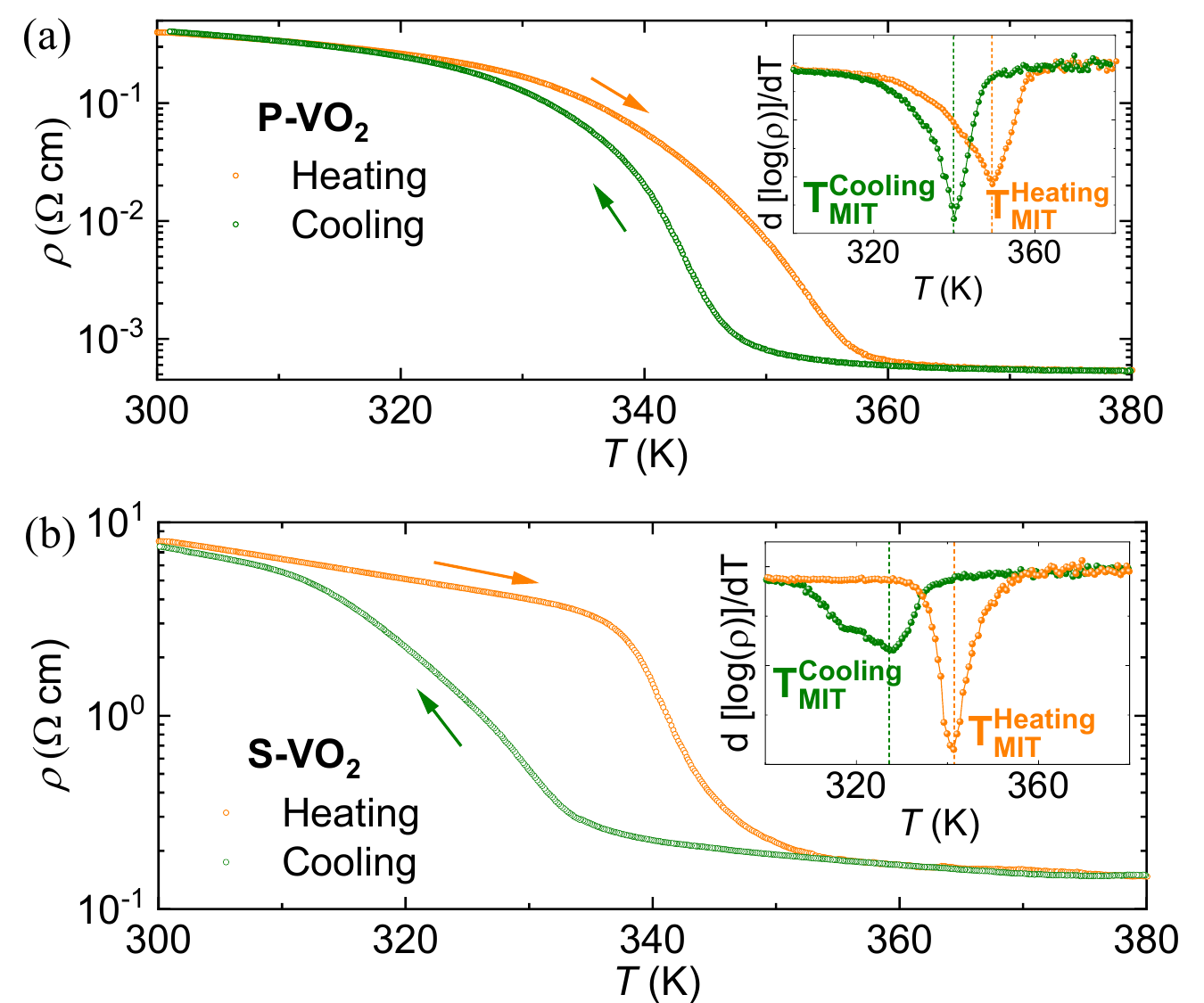}
		\caption{Resistivity vs temperature curve of (a) P-VO$_2$ and (b) S-VO$_2$ thin film. Inset figure shows the temperature derivative of log$(\rho)$.}
		\label{fig1}
	\end{center}
\end{figure}

Figures \ref{fig1}\textcolor{blue}{a} and \ref{fig1}\textcolor{blue}{b} display the resistivity vs temperature ($\rho$ $-T$) for P-VO$_2$ and S-VO$_2$, respectively. P-VO$_2$ exhibits a MIT upon heating at 348 K with three orders of magnitude change in resistivity and a thermal  hysteresis of about 9 K. In contrast, S-VO$_2$ shows  MIT at 341 K with only two orders of magnitude change in resistivity and relatively larger thermal hysteresis of about 16 K. Such a large hysteresis could be linked to broad metal-insulator domain distribution near $T_\mathrm{MIT}$ \cite{fisher2017switching}. The cause of a wider transition in S-VO$_2$ is complex and it depends on several factors such as non-uniform distribution of different grains, strain and oxygen stoichiometry. \cite{morin1959oxides,mott1968metal,morrison2014photoinduced,qazilbash2007mott,muraoka2014persistent, yang2015surface, muraoka2002metal}. The shorter b-axis may be attributed to the presence of oxygen deficiency that possibly stabilize the metallic domains over a wider temperature range, enhancing phase coexistence and broadening the hysteresis loop \cite{zhang2017evolution}.
As shown in the insets of Fig. \ref{fig1}, we defined the MIT temperatures from the peaks of the derivative dlog($\rho$)/dT. During heating and cooling, P-VO$_2$ exhibits symmetric behavior in resistivity, while S-VO$_2$ displays an asymmetric nature, especially on cooling (Fig \ref{fig1}\textcolor{blue}{b}).

\begin{figure*}
	\begin{center}
		\includegraphics[scale=0.5]{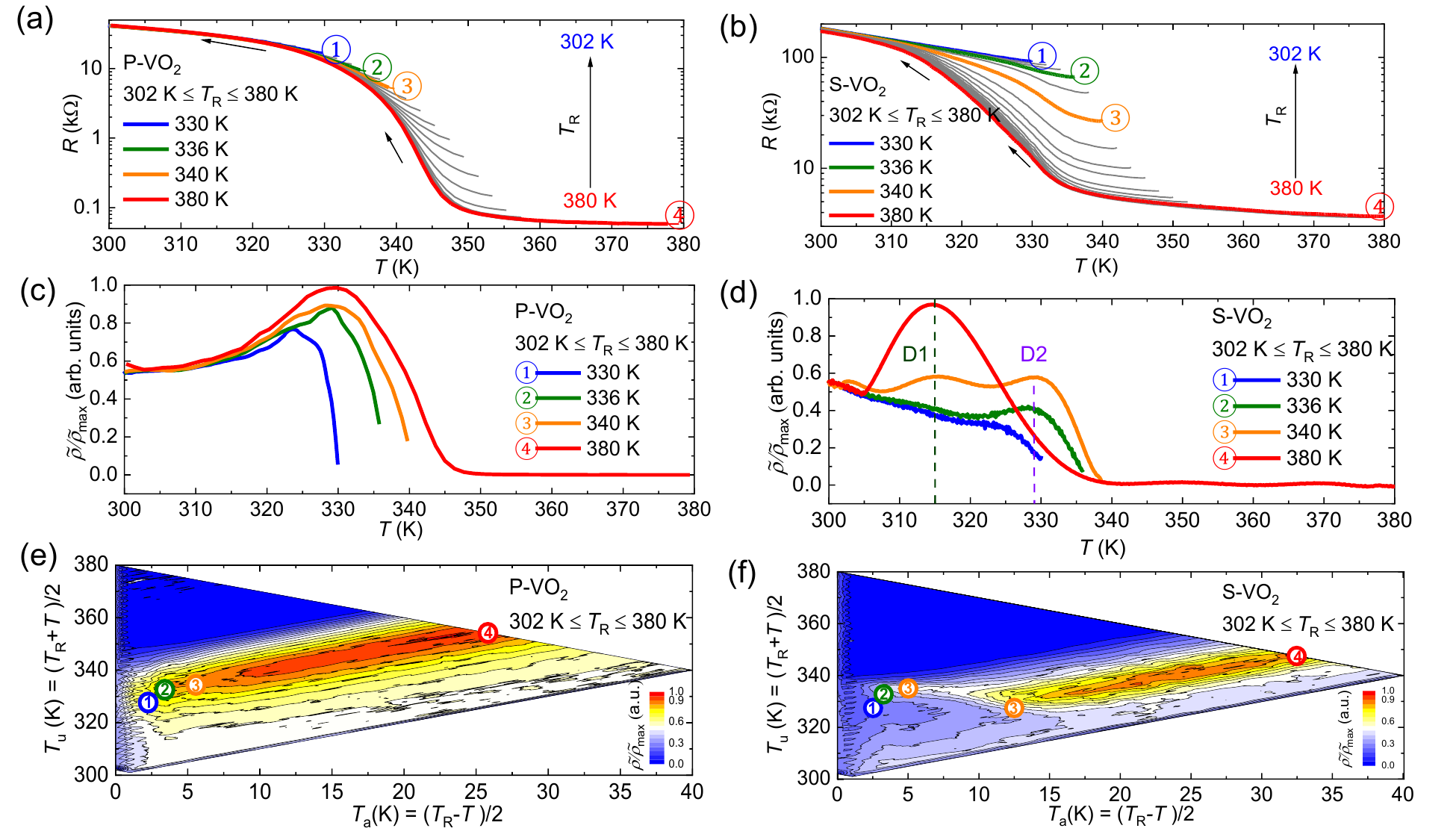}
		\caption{The family of cooling FORCs for (a) P-VO$_2$ and (b) S-VO$_2$ films. (c-d) Out of this forty FORCs (shown in grey), four curves 1-4 are selected in color. FORC distribution function is plotted against temperature for these two samples. (e-f) The FORC distribution contour plot with new coordinate ($T_\mathrm{a}$, $T_\mathrm{u}$) as defined in the main text. Blue and red color reflect the reversible and irreversible nature of the phase, respectively.}
		\label{fig2}
	\end{center}
\end{figure*}

To obtain a detailed microscopic perspective of the phase evolution, and gain information regarding the distribution of domains and their interactions, we exploit the FORC method \cite{ramirez2009first} (supplementary for details of measurement process). Figs. \ref{fig2}\textcolor{blue}{a} and \ref{fig2}\textcolor{blue}{b} display the cooling FORC data obtained for P-VO$_2$ and S-VO$_2$, respectively. A total of 40 reversal curves were recorded for each sample. For illustrative purposes, we select four curves.

The FORC distribution is determined by evaluating the mixed second-order derivative \cite{ramirez2009first,mayergoyz1991vector,katzgraber2002reversal,pike1999characterizing,roberts2014understanding,frampton2017first}
\begin{equation}
	\tilde{\rho }\left(T_\mathrm{R},T\right)=- \frac{1}{2} \frac{\partial^2 R\left(T_\mathrm{R},T\right)}{\partial T_\mathrm{R} \partial T}
\end{equation}
where $R$ ($T_\mathrm{R}$,$T$) corresponds to the resistance of the film. The derivative $\partial R$($T, T_\mathrm{R}$) /$\partial T$ represents the slope of the function $R$ ($T_\mathrm{R}$,$T$) at various values of $T$. Similarly, the derivative $\partial$$T_\mathrm{R}$ captures how the slope changes at a specific $T$ value along different branches originating from various $T_\mathrm{R}$ values. The mixed second-order derivative disregards the data set for which the change in resistance with respect to $T$ or \textit{T$_\mathrm{R}$} is constant. Consequently, any non-zero value within the distribution indicates the presence of irreversible regions within the hysteresis loop \cite{mayergoyz1991vector,pike1999characterizing,roberts2014understanding}. Figures \ref{fig2}\textcolor{blue}{c} and \ref{fig2}\textcolor{blue}{d} display the FORC distributions ($\tilde{\rho }$) of P-VO$_2$ and S-VO$_2$, respectively. In case of P-VO$_2$, a peak at about 330 K is noticed. This single peak indicates a uniform distribution of domains, which undergo a metal-insulator domain switching at 330 K. Interestingly, for S-VO$_2$  $\tilde{\rho }$ exhibits two distinct peaks (Fig. \ref{fig2}\textcolor{blue}{d}): a dominant peak at $T_\mathrm{{p_1}}$$\sim$ 316 K and a secondary peak at $T_\mathrm{{p_2}} $$\sim$ 331 K. These two peaks signify two distinct domain distributions (marked as D1 and D2), extending over 305$-$338 K and 342$-$380 K respectively. 

The presence of two peaks in the FORC distribution of S-VO$_2$ is consistent with models that describe supercooled metallic domains stabilised by inter-grain interactions. In this framework, each grain has a distinct critical temperature, but an interaction-induced energy barrier can delay switching, allowing some metallic regions to persist at lower temperatures. These domains can act as seeds during reversal, altering the phase transition path \cite{ramirez2009first,kim2022nanoscale,sohn2015fractal,gu2022random}.
This behavior is further observed by Raman spectroscopy (Fig.\textcolor{blue}{S2}). Raman analysis shows that P-VO$_2$ undergoes a direct rutile-to-monoclinic transition during cooling, while S-VO$_2$ exhibits an intermediate state. The smaller grains in S-VO$_2$ promote the stabilization of a metastable state, where supercooled domains form in the insulating matrix \cite{guo2011mechanics,hattori2020investigation,rodriguez2021strong}. This results in multi-directional percolation and an asymmetric hysteresis, as reflected in the FORC distribution.

A coordinate transformation from  $\tilde{\rho }$\{$T_\mathrm{R}$, $T$\} to $\tilde{\rho }$\{$T_\mathrm{a}$, $T_\mathrm{u}$\} is carried out for understanding the phase-specific transition temperatures and thermal hysteresis linked with the activation energy. Here, $T_\mathrm{a}$ = $\frac{(T_\mathrm{R}-T)}{2}$, physically corresponds to the energy barrier or activation energy associated with the phase transition, while $T_\mathrm{u}$ = $\frac{(T_\mathrm{R}+T)}{2}$ represents the average temperature of the phase transition \cite{frampton2017first,gilbert2021reconstructing}. Figures \ref{fig2}\textcolor{blue}{e} and \ref{fig2}\textcolor{blue}{f} show $\tilde{\rho }$ in the 2D color contour plot. In this plot, blue and red colors represent the reversible and irreversible nature of the phase, respectively. In case of P-VO$_2$, the irreversible nature ($\tilde{\rho }$ = 0) is observed below $T_\mathrm{u}$ $\sim$ 345 K whereas in S-VO$_2$ the irreversibility is noticed below $T_\mathrm{u}$ $\sim$ 338 K. The irreversibility pattern in the P-VO$_2$ system is unidirectional as visualized by connecting the 4 points (the encircled numbers in Fig. \ref{fig2}\textcolor{blue}{e}), suggesting a single conduction channel with a spread along $T_\mathrm{u}$ and an elongated continuous pattern along the $T_\mathrm{a}$ direction. Such unidirectional irreversibility in P-VO$_2$ associated with the metal-insulator transition (MIT) consistent with previous literature \cite{sharoni2008multiple,ramirez2009first,sohn2015fractal,kim2022nanoscale}. In contrast, S-VO$_2$ displays a bidirectional irreversibility pattern, revealing the presence of two distinct conduction channels with two different spreads along $T_\mathrm{u}$ and an elongated discrete pattern along the $T_\mathrm{a}$ direction. Such bidirectional irreversibility is attributed to supercooled metallic domains\cite{ramirez2009first,davies2004magnetization}. These metallic domains act as nucleation centers for the transition, modifying the MIT pathway through interactions with the neighboring insulating matrix, which is also a possible reason for the asymmetric thermal hysteresis. The origin behind such distribution may be related to the growth processes (crystallite size, strain, oxygen vacancy, defects etc. \cite{kim1994pulsed,jeong2013suppression,brassard2005grain,youn2004growth}) If every crystallite possesses identical properties, it would show the same activation energy, resulting in common $T_\mathrm{a}$ \cite{frampton2017first}. The formation of crystallites effects  the common conduction channel as well as the domain nucleation. The P-VO$_2$ film has a uniform distibution of crystallite (continuous red patches in Fig. \ref{fig2}\textcolor{blue}{e}) with a common $T_\mathrm{a}$ that effectively results in a direct transition from R to M1 phase. On the contrary, S-VO$_2$ possesses shorter $b$-axis which results metastable phases due to non-uniform crystallite distribution (discrete red region as shown in Fig. \ref{fig2}\textcolor{blue}{f}).  \par
\begin{figure*}
	\begin{center}
		\includegraphics[scale=0.6]{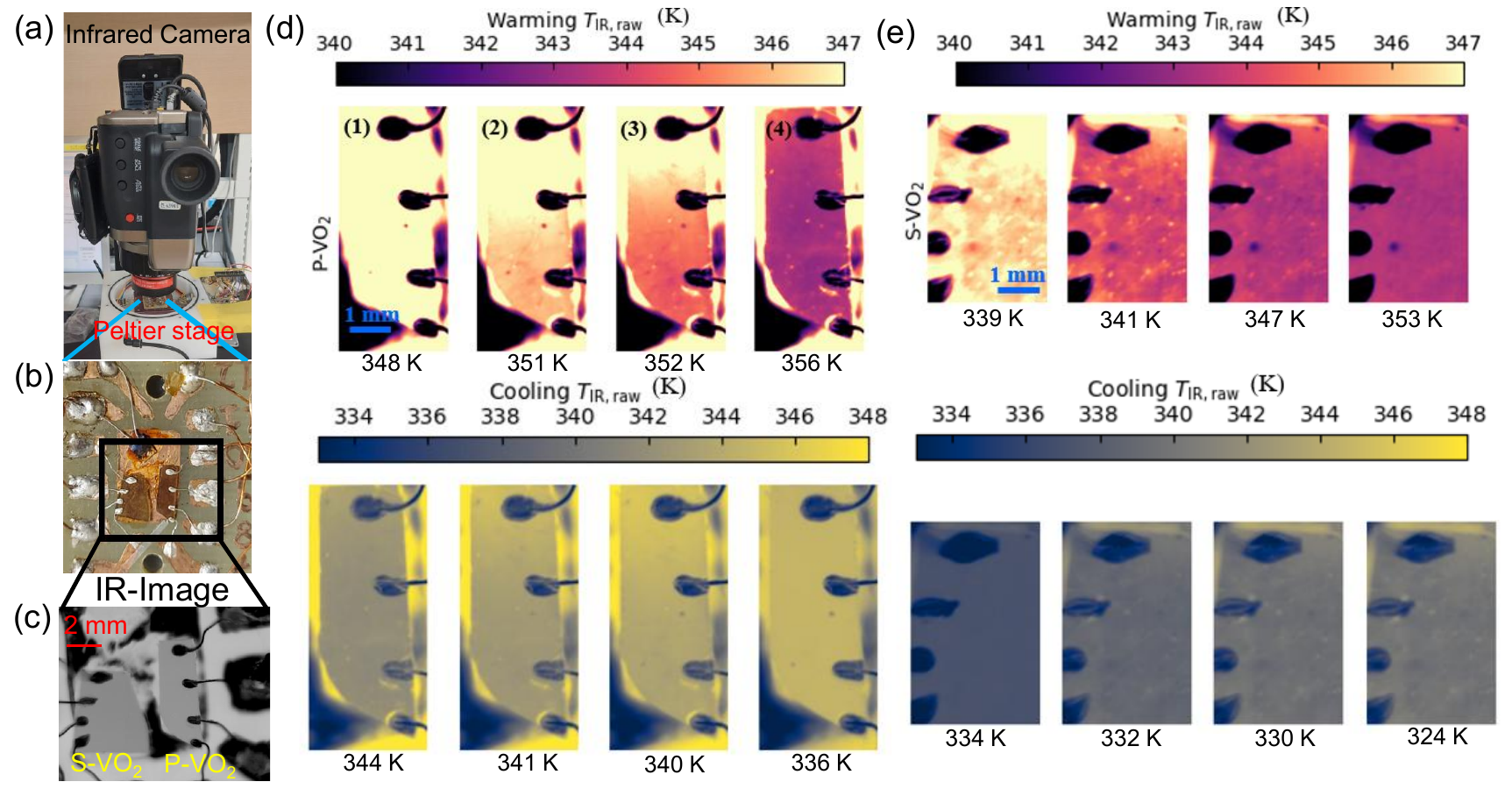}
		\caption{(a) Photograph of thermal imaging set-up which consisting of IR camera and Peltier stage, (b) sample holder with copper pads, and (c) IR image recorded at room temperature, where the black region on top of the samples are the Ag paint that used to attach the contact wires. Temperature-driven thermal imaging across the MIT of (d) P-VO$_2$ and (e) S-VO$_2$ during heating and cooling. The color scale indicates the raw values of local temperature measured by infrared camera. Due to higher reflectivity, metallic regions appear cooler (darker), while the insulating region, with lower reflectivity, appears warmer (lighter).}
		\label{fig3}
	\end{center}
\end{figure*}
In addition to the macroscopic phase evolution, the hysteretic behavior of VO$_2$ also depends on processes of metal-insulator domain nucleation and growth \cite{choi1996mid,mcleod2017nanotextured,qazilbash2009infrared,qazilbash2007mott,qazilbash2007mott}. To obtain a microscopic view of these processes, we performed IR imaging. Figure \ref{fig3}\textcolor{blue}{a} shows the thermal imaging set-up that we used to capture the optical maps of the raw IR temperature ($T_\mathrm{raw}$). The sample holder consists of a FR-4 glass epoxy plate (thickness 1.6 $\mu$m), covered with a thin copper layer (35 $\mu$m) (Fig. \ref{fig3}\textcolor{blue}{b}). GE varnish was used to attach the sample on the holder. The resistance of VO$_2$ is used as a self-consistent thermometer by using its resistance vs temperature curve measured by PPMS. The temperature re-scaling procedure is described in the supplementary Fig.\textcolor{blue}{S4}. Figure \ref{fig3}\textcolor{blue}{c} shows a typical IR image captured at room temperature. Thermal images on VO$_2$ films have been recorded during both heating and cooling temperature ramps (0.87 K/min). Figures \ref{fig3}\textcolor{blue}{d} and \ref{fig3}\textcolor{blue}{e} show a subset of thermal images of P-VO$_2$ and S-VO$_2$. IR camera captures the local temperature distribution of the sample and allows to distinguish metallic and insulating regions from their different emissivity. The IR images are presented with a colour scale indicating the raw value of thermal temperature ($T_\mathrm{{raw}}$) measured by the infrared camera with an emissivity of 1, as described in \cite{mattoni2020role}.
To ensure accurate thermal contrast, we performed separate IR calibrations for P-VO$_2$ and S-VO$_2$. Since each sample exhibits uniform surface roughness and grain size across the field of view, the emissivity is constant within a given sample and phase, so that the observed thermal contrast corresponds primarily to phase changes. 
The change in the emissivity of VO$_2$ across the phase transition allows us to capture the progressive evolution and expansion of the metallic phase upon heating. Since the reflectivity of the metallic regions is larger, metallic regions appear colder (i.e., darker) than insulating regions that appear hotter (i.e., brighter) in $T_{\mathrm{raw}}$.
To clearly visualise the features across the MIT, we used two different color scales of 340–347 K for heating and 334–348 K for cooling. We have checked the IR temperature of the sample holder in six different places, and we found the IR temperature of the sample holder to be constant over a distance of 10 mm, thus ensuring homogeneous temperature distribution as shown in Fig.\textcolor{blue}{S4}.

In the heating cycle shown in Figs. \ref{fig3}\textcolor{blue}{d} and \ref{fig3}\textcolor{blue}{e}, a small metallic fragment emerges and rapidly propagates throughout the film, indicating a progressive transition of insulating to metallic domains.  Metallic domains nucleate as droplets within the crystallite of VO$_2$ and evolve into clusters of metallic phase with increase in $T$. As $T$ rises above $T_\mathrm{MIT}$, the conducting clusters grow in size and coalesce, eventually leading to a conducting channel \cite{choi1996mid}. 
In P-VO$_2$, we see nucleation and growth from the bottom side of the image (Fig. \ref{fig3}\textcolor{blue}{d}), possibly due to larger crystallites determining a spatially continuous MIT. Larger domains may suppress grain-boundary nucleation, causing the transition to propagate continuously along specific crystallographic directions \cite{liu2015phase}.
\begin{figure*}
	\begin{center}
        \includegraphics[scale=0.5]{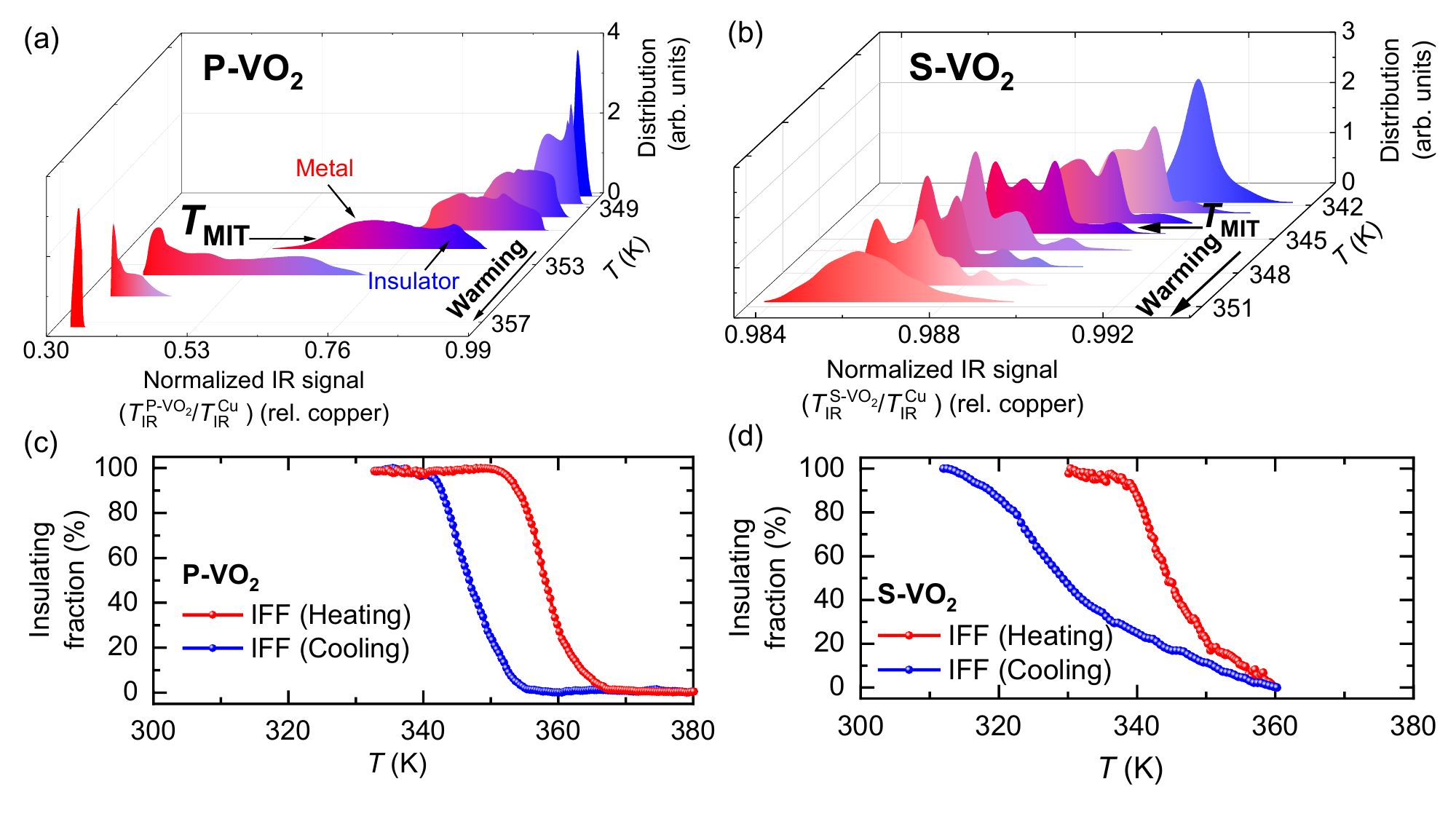}
		\caption{ Quantitative analysis of IR-imaging across the MIT: (a),(b) histogram of local sample temperatures relative to that of copper, collected from the films during heating from 330 K to 380 K. A uniform distribution for P-VO$_2$ and a multimodal distribution for S-VO$_2$. The insulating fill fractions (IFF) with temperature for (c) P-VO$_2$ and (d) S-VO$_2$.}
		\label{fig4}
	\end{center}
\end{figure*}
In S-VO$_2$, instead, the nucleation is rather sparse over the sample surface (as shown in Fig. \ref{fig3}\textcolor{blue}{e}). These channels are consistent with our assessment of a unidirectional conduction channel in P-VO$_2$, and a multidirectional channel in S-VO$_2$ \cite{mcleod2017nanotextured}. The IR images demonstrate the insulator-metal transition in four consecutive states on heating: (1) a uniform insulating state, (2) a non-uniform insulating state with isolated metallic fragments, (3) a phase-coexistence state, and finally (4) a uniform metallic state \cite{choi1996mid,mcleod2017nanotextured,qazilbash2009infrared,mcleod2017nanotextured,imada1998metal}. Conversely, upon cooling, these metallic domains became disconnected from one another, leading to the formation of disordered patches that eventually dissolved back into insulating domains.
During heating and cooling, the metallic and insulating domains in P-VO$_2$ evolve and grow continuously from one side of the sample, whereas in S-VO$_2$ they nucleate sparsely, possibly influenced by supercooled metallic domains as shown in supplementary Fig. \textcolor{blue}{S6}. The presence of these domains leads to an asymmetry in the MIT hysteresis in S-VO$_2$.
Although a quantitative estimation of metallic fragment size is beyond our instrumental resolution limit, a clear difference between the two samples is evident.

For a quantitative comparison, the intensity of the IR signal was normalized with that of the copper pads on the sample holder, assuming the emissivity of the copper to be constant throughout the temperature ramp (Fig. \textcolor{blue}{${\mathrm{S5}}$}).
We here assume that the infrared temperature of the material `i' (i = copper pad of the sample holder, VO$_2$ metal, or VO$_2$ insulator) can be expressed as
\begin{equation}
	T_\mathrm{IR}^\mathrm{i} = A_\mathrm{i} + B_\mathrm{i} T_\mathrm{raw} 
\end{equation}
where $A_\mathrm{i}$, $B_\mathrm{i}$ are material-specific constants \cite{mattoni2020role}. For the copper layer, we assume $A_\mathrm{i}$ and $B_\mathrm{i}$ to be temperature independent, while for VO$_2$ two different values need to be used for the metallic and the insulating phase. After our normalization, the color of the metallic and insulating phases become temperature-independent, as described by the following relation:
\begin{equation}
\mathrm{\left[{color\ of \atop metallic/insulating} \right]} = \frac{\left[T_\mathrm{IR,raw} - T_\mathrm{IR,raw, Cu}\mathrm{(}T=\mathrm{317 K)}\right] }{ \left[T_\mathrm{IR,raw,Cu} - T_\mathrm{IR,raw,Cu} \mathrm{(}T=\mathrm{317 K)}\right]}.
\end{equation}
\noindent Following this normalization, the color of the metallic and insulating phases are represented in the histograms in Figs. \ref{fig4}\textcolor{blue}{a} and \ref{fig4}\textcolor{blue}{b} by red and blue, respectively.

The 3D histogram color plots are used to visualize the distribution of IR intensities across the MIT. It illustrates the binary characteristic of first-order phase transition \cite{qazilbash2009infrared,mcleod2017nanotextured} at various temperatures during the heating. Each bimodal distribution represents two distinct populations (insulating: blue region, metallic: red region). For P-VO$_2$ shown in Fig. \ref{fig4}\textcolor{blue}{a}, the metallic population evolves with temperature, and a single peak evolves into a double peak through the transition, then becomes a single peak for the metallic phase.\cite{frenzel2018infrared,qazilbash2009infrared,mcleod2017nanotextured} 
Such a bimodal distribution at the MIT suggests a direct transition from R to M1 phase, consistent with the FORC analysis. In contrast, S-VO$_2$ shows multi-distribution through the MIT (Fig. \ref{fig4}\textcolor{blue}{b}), indicating that metallic and insulating domains with different characteristics may form, consistent with multiple conduction channels identified by our FORC analysis.
\begin{table}
	\small
	\caption{{Comparison of P-VO$_2$ and S-VO$_2$ film}}
	\label{tbl5}
	\begin{tabular*}{0.48\textwidth}{@{\extracolsep{\fill}}lllllllllll}
		\hline
		\hline
		Parameters &   P-VO$_2$  &    S-VO$_2$             \\ \hline
		$b$-axis lattice parameter & 4.53 \AA&4.41 \AA\\
		RMS roughness (nm) &0.53 &4.27 \\
		Growth pattern & layer-by-layer& island-like\\
		$T_\mathrm{MIT}$ (Heating) & 348 K&    341 K       \\
		Hysteresis width  [($\Delta T) _{\rho-T}$]  & 9 K& 16 K\\
		Irreversibility pattern & unidirectional& bidirectional\\
		Domain distribution & uniform &  non-uniform\\
		Domain percolation & single conduction \\&channel& multiple channel\\
		Threshold percolation \\ temperature (heating) &$\sim$ 356 K & $\sim$ 342 K\\
		Hysteresis nature  &symmetric & asymmetric\\	
		\hline
		\hline
	\end{tabular*}
\end{table}

We plot the insulating fill fraction (IFF) that is the percentage of insulating area estimated using the change of emissivity \cite{mcleod2017nanotextured} as a function of temperature in Fig. \ref{fig4}\textcolor{blue}{c} and \ref{fig4}\textcolor{blue}{d} for both films. During heating the threshold percolation temperature (50\% of IFF) is 356 K for P-VO$_2$ and 342 K for S-VO$_2$. The IFF from thermal image analysis matches with estimated insulating volume fraction from the resistivity measurement for P-VO$_2$ and S-VO$_2$. Surface analysis of P-VO$_2$ shows symmetric hysteretic behaviour during heating and cooling whereas S-VO$_2$ exhibits a large asymmetry similar to what measured by resistivity. During cooling, $T_\mathrm{MIT}$ of S-VO$_2$ is broad probably due to interfacial compressive strain arising from supercooled metallic domains that prevent structural relaxation \cite{yang2015surface,del2018resistive,fan2011large}. A comparative table of structure and transport parameters are summarized in table \ref{tbl5}.

In conclusion, the combined use of FORC and IR imaging allowed us to study the MIT in VO$_2$ thin films of distinct grain size.
In larger-grain samples, the MIT is continuous characterized by a single conduction channel and a symmetric thermal hysteresis. In smaller-grain samples, instead, the MIT is characterized by sparse domains, multiple conduction channels, and asymmetric hysteresis, consistent with the presence of supercooled metallic domains.
These contrasting behaviors underscore how growth-induced microstructural variations govern domain percolation and hysteretic response in VO$_2$.

\subsection*{Supplementary material}
See the supplementary material for details of the structural characterization, Raman spectroscopy analysis, FORC measurement procedures, IR image analysis, and domain boundary across MIT in VO$_2$ thin films.

\subsection*{ Acknowledgments}
We are grateful to Yoshiteru Maeno for useful discussions. C.S.
acknowledges research support from IIT Kanpur Initiation Grant
(IITK-2019-037) and research grants from the Science and
Engineering Research Board (SERB), Government of India (Grant Nos.
SRG2019-001104, CRG-2022-005726, and EEQ-2022-000883). A.V.
acknowledges a research grant from Science and Engineering Research
Board (SERB) (Grant No. CRG/2022/005421).

\subsection*{AUTHOR DECLARATIONS}
\subsection*{Conflict of Interest}
The authors declare no competing interests.
\subsection*{Author Contributions}
\textbf{Shubhankar Paul:} Conceptualization (equal); Data curation (lead); Formal analysis (lead); Investigation (lead); Methodology (lead); Validation (equal); Visualization (equal); Writing – original draft (lead); Writing – review \& editing (equal)
\textbf{Giordano Mattoni}: Data curation (supporting); Formal analysis (equal); Investigation (equal); Methodology (equal); Validation (equal); Writing - original draft (equal).
\textbf{Amitava Ghosh:}  Investigation (supporting); Methodology (supporting).
\textbf{Pooja Kesarwani:} Investigation (supporting); Methodology (supporting).
\textbf{Dipak Sahu:} Investigation (supporting); Methodology (supporting).
\textbf{Monika Ahlawat:}Investigation (supporting); Methodology (supporting).
\textbf{Ashok P:}  Investigation (supporting); Methodology (supporting).
\textbf{Amit Verma:} Funding acquisition (supporting);
Investigation (supporting); Methodology (supporting); Project administration (supporting); Resources (supporting); Validation (supporting);
Writing – original draft (supporting). 
\textbf{Vishal Govind Rao} Funding acquisition (supporting);
Investigation (supporting); Methodology (supporting); Project administration (supporting); Resources (supporting); Validation (supporting);
Writing – original draft (supporting). 
\textbf{Chanchal Sow:}  Conceptualization
(equal); Formal analysis (equal); Funding acquisition (lead); Investigation
(equal); Project administration (lead); Resources (lead); Supervision
(lead); Validation (equal); Visualization (equal); Writing – original draft
(equal); Writing – review \& editing (equal).
\subsection*{DATA AVAILABILITY}
The data that support the findings of this study are available from
the corresponding authors upon reasonable request.

\bibliography{ref_resub}
\bibliographystyle{apsrev4-2}
\end{document}